\documentclass[preprint]{seismica}

\usepackage{subcaption}

\title{Pervasive Label Errors in Seismological Machine Learning Datasets}

\author[1]{Albert Leonardo Aguilar Suarez
	\orcid{0000-0002-6496-3450}
	\thanks{Corresponding author: aguilars@stanford.edu}
}
\author[1]{Gregory C. Beroza
	\orcid{0000-0002-8667-1838}
}

\affil[1]{Department of Geophysics, Stanford University, Stanford, California}

\credit{Conceptualization}{Aguilar Suarez, Beroza}
\credit{Methodology}{Aguilar Suarez}
\credit{Formal Analysis}{Aguilar Suarez, Beroza}
\credit{Writing - Original draft}{Aguilar Suarez, Beroza}

\begin{document}

	
	\makeseistitle{
		\begin{summary}{Abstract}
			The recent boom in artificial intelligence and machine learning has been powered by large datasets with accurate labels, combined with algorithmic advances and efficient computing. 
            The quality of data can be a major factor in determining model performance. Here, we detail observations of commonly occurring errors in popular seismological machine learning datasets. We used an ensemble of available deep learning models PhaseNet and EQTransformer to evaluate the dataset labels and found four types of errors ranked from most prevalent to least prevalent: (1) unlabeled earthquakes; (2) noise samples that contain earthquakes; (3) inaccurately labeled arrival times, and (4) absent earthquake signals. We checked a total of 8.6 million examples from the following datasets: Iquique, ETHZ, PNW, TXED, STEAD, INSTANCE, AQ2009, and CEED. The average error rate across all datasets is 3.9 \%, ranging from nearly zero to 8 \% for individual datasets. These faulty data and labels are likely to degrade model training and performance. By flagging these errors, we aim to increase the quality of the data used to train machine learning models, especially for the measurement of arrival times, and thereby to improve the reliability of the models. We present a companion list of examples that contain problems, aiming to integrate them into training routines so that only the reliable data is used for training.
		\end{summary}
	    }  
	

	\section{Introduction}
    Data and algorithms are the pillars of the recent explosion in machine learning implementations for seismology \citep{arrowsmith22, dlseismo}. The volume, variety, and veracity of data available to seismologists have fueled many deep learning models for the measurement of arrival times, which in turn have fueled a new generation of earthquake catalogs that contain many more earthquakes than previously known in routine catalogs \citep{Tan2021,Schultz2023,tan2025}. The landscape of datasets for observational seismology is dominated by local earthquake recordings from densely instrumented regions \citep{stead,instance,zhu2025californiaearthquakedatasetmachine}.
    There is a growing number of datasets that aim to fill the gaps in other observational settings, including seismograms from: regional distances \citep{AguilarSuarez_Beroza_2024,PnSn}, moderate magnitude earthquakes \citep{Tang2024}, ocean bottom seismometers \citep{Bornstein2024}, magmatic systems \citep{Zhong2024}, and induced seismicity \citep{texed}. These datasets are the basis for many models \citep{Zhu2019,EQTransformer,Niksejel2024,AguilarSuarez_Beroza_2025} that are trained to identify and measure the arrival times of seismic waves. Arrival time measurements are the basis for the downstream tasks of phase association, location, and magnitude estimation, for which there are also deep learning implementations.

    \begin{figure*}

        \begin{subfigure}[b]{0.5\textwidth}
        \centering
        \includegraphics[trim={2.5cm 0cm 2.5cm 0cm},clip,width=\textwidth]{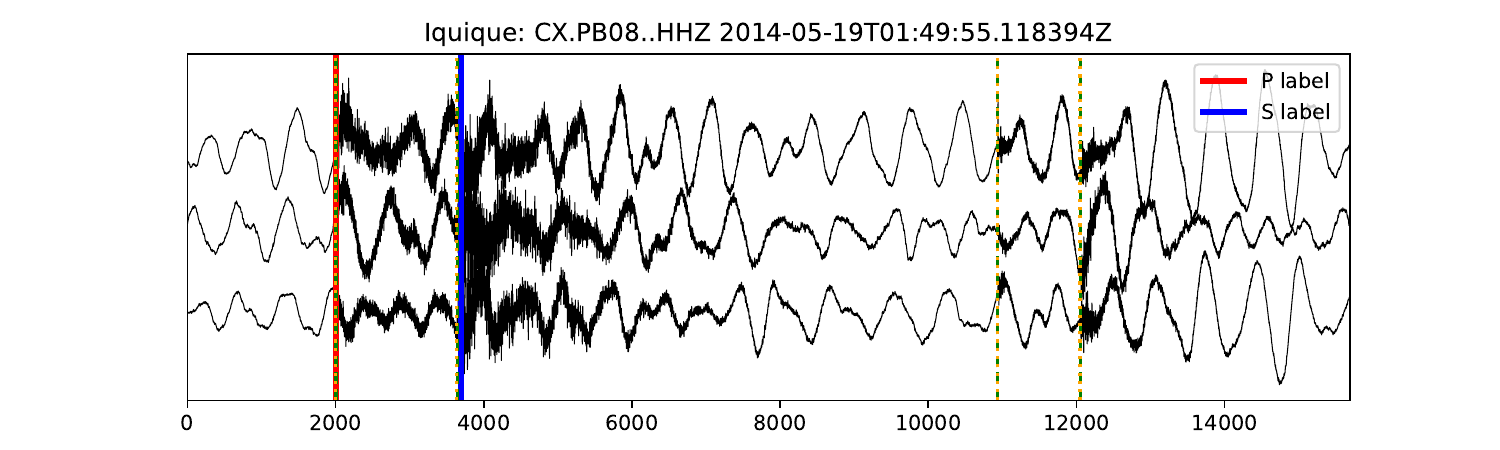}
        \end{subfigure}
        \hfill
        \begin{subfigure}[b]{0.5\textwidth}
        \centering
        \includegraphics[trim={2.5cm 0cm 2.5cm 0cm},clip,width=\textwidth]{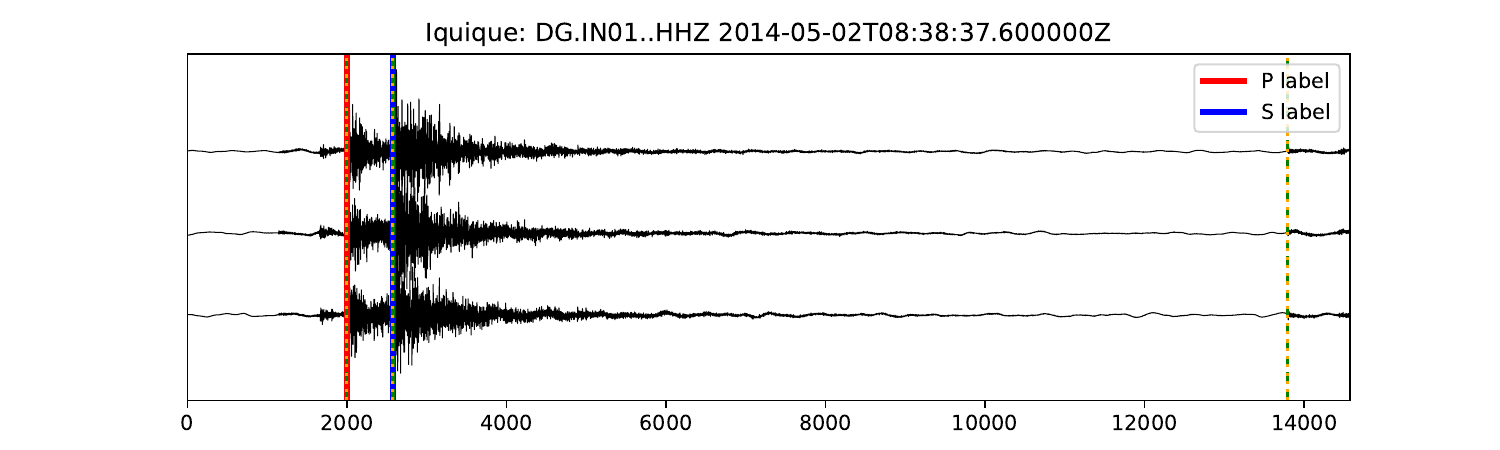}
        \end{subfigure}

                \begin{subfigure}[b]{0.5\textwidth}
        \centering
        \includegraphics[trim={2.5cm 0cm 2.5cm 0cm},clip,width=\textwidth]{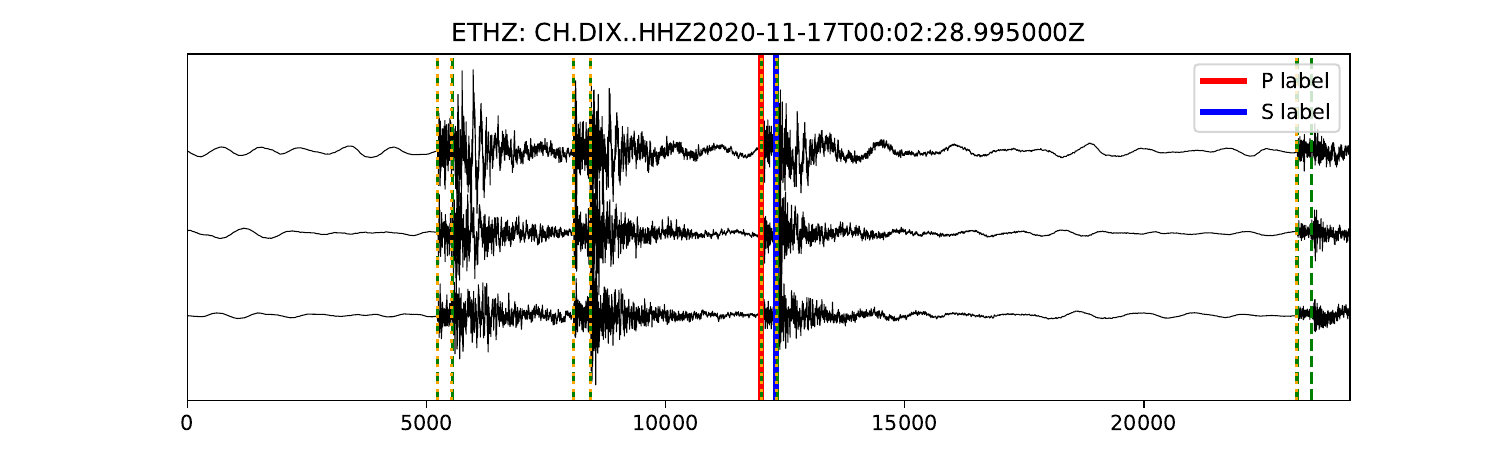}
        \end{subfigure}
        \hfill
        \begin{subfigure}[b]{0.5\textwidth}
        \centering
        \includegraphics[trim={2.5cm 0cm 2.5cm 0cm},clip,width=\textwidth]{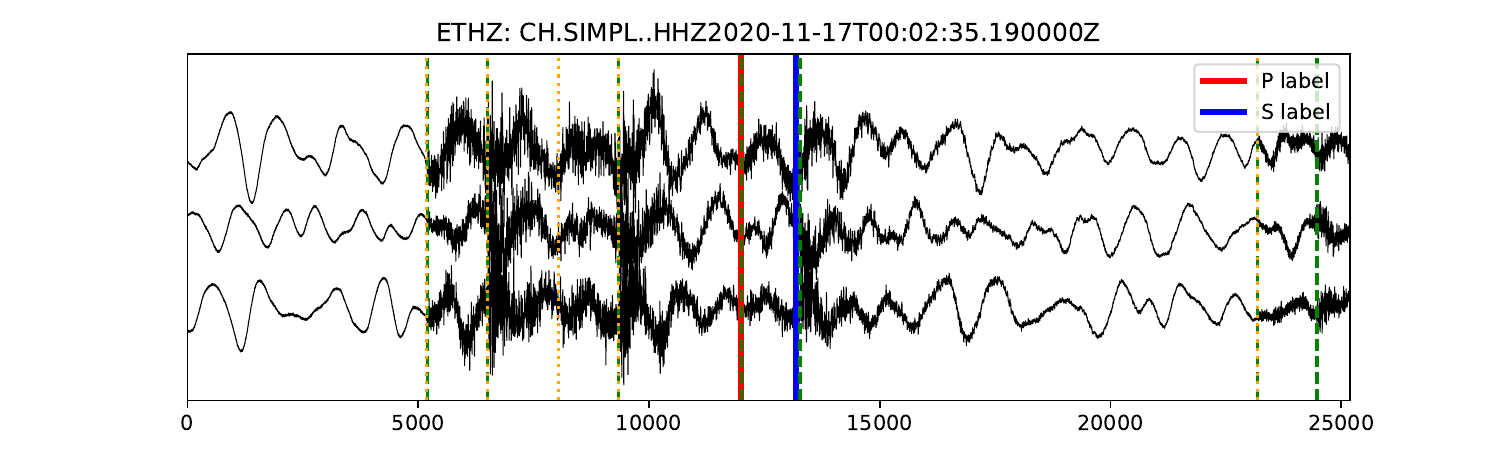}
        \end{subfigure}

        \begin{subfigure}[b]{0.5\textwidth}
        \centering
        \includegraphics[trim={2.5cm 0cm 2.5cm 0cm},clip,width=\textwidth]{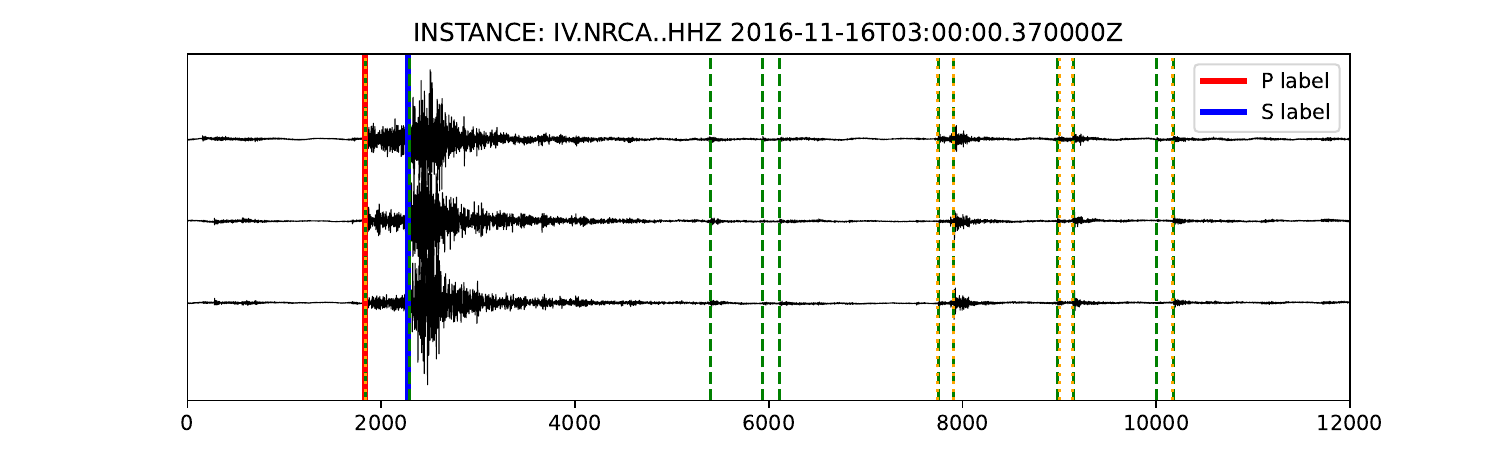}
        \end{subfigure}
        \hfill
        \begin{subfigure}[b]{0.5\textwidth}
        \centering
        \includegraphics[trim={2.5cm 0cm 2.5cm 0cm},clip,width=\textwidth]{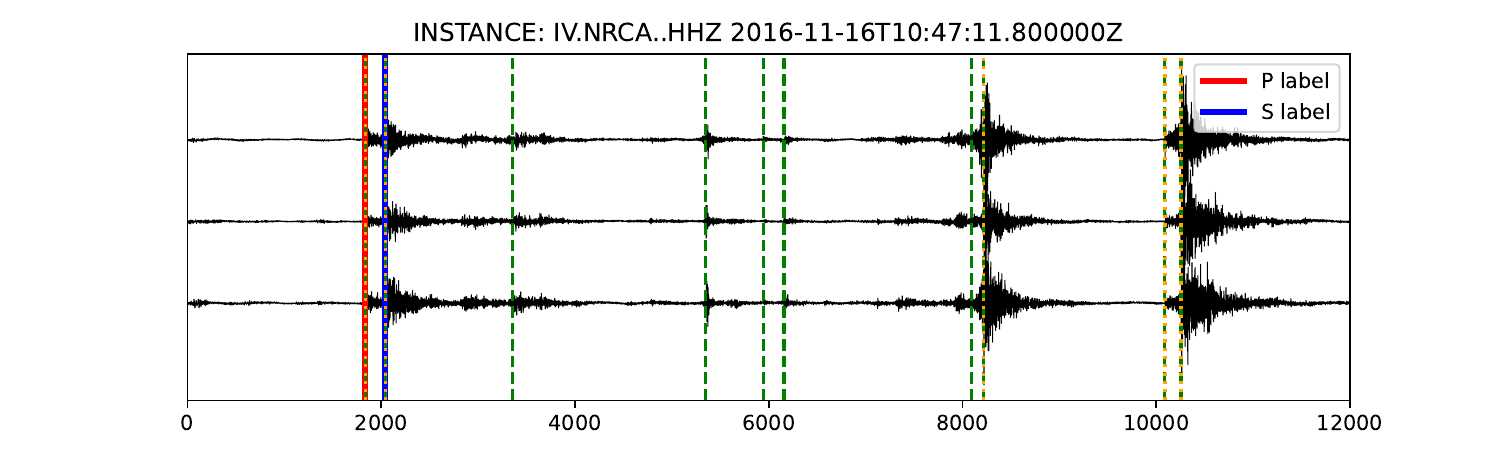}
        \end{subfigure}
        
        \begin{subfigure}[b]{0.5\textwidth}
        \centering
        \includegraphics[trim={2.5cm 0cm 2.5cm 0cm},clip,width=\textwidth]{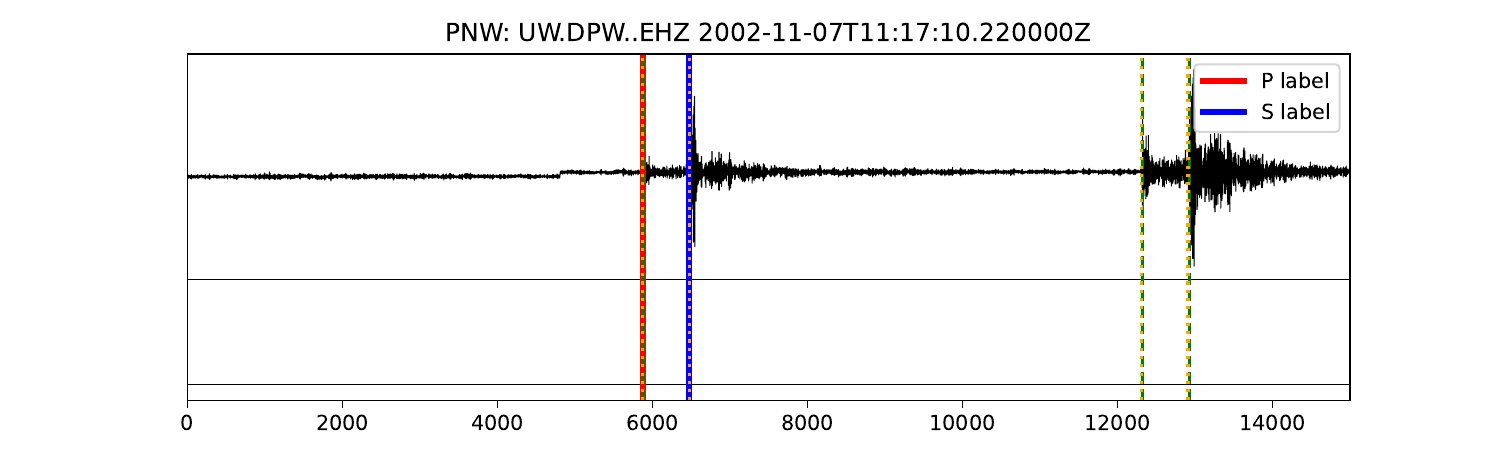}
        \end{subfigure}
        \hfill
        \begin{subfigure}[b]{0.5\textwidth}
        \centering
        \includegraphics[trim={2.5cm 0cm 2.5cm 0cm},clip,width=\textwidth]{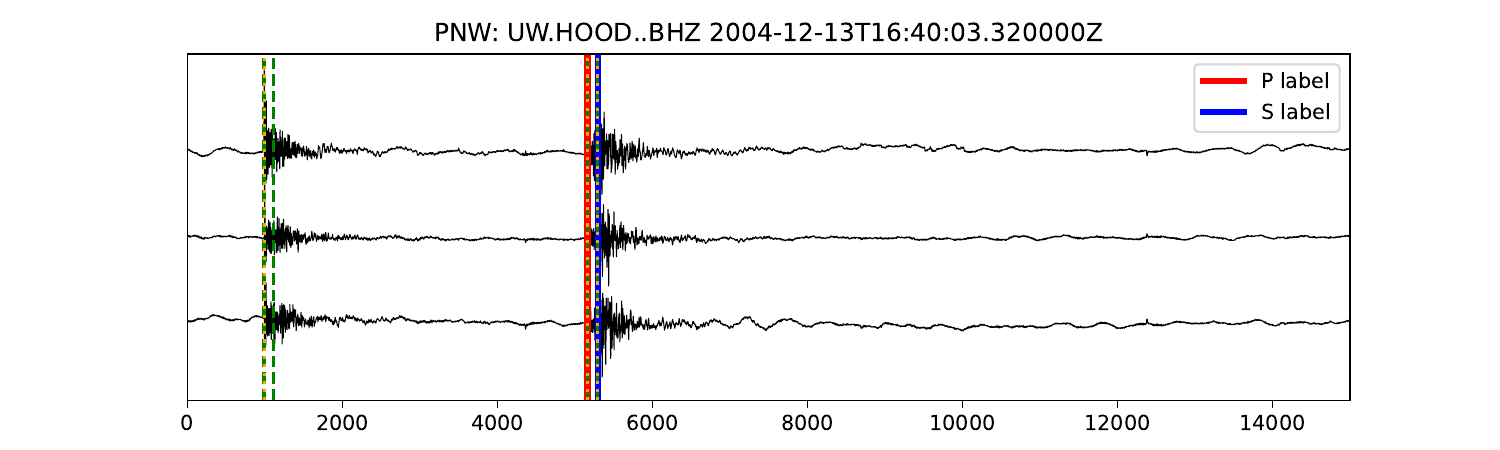}
        \end{subfigure}

        \begin{subfigure}[b]{0.5\textwidth}
        \centering
        \includegraphics[trim={2.5cm 0cm 2.5cm 0cm},clip,width=\textwidth]{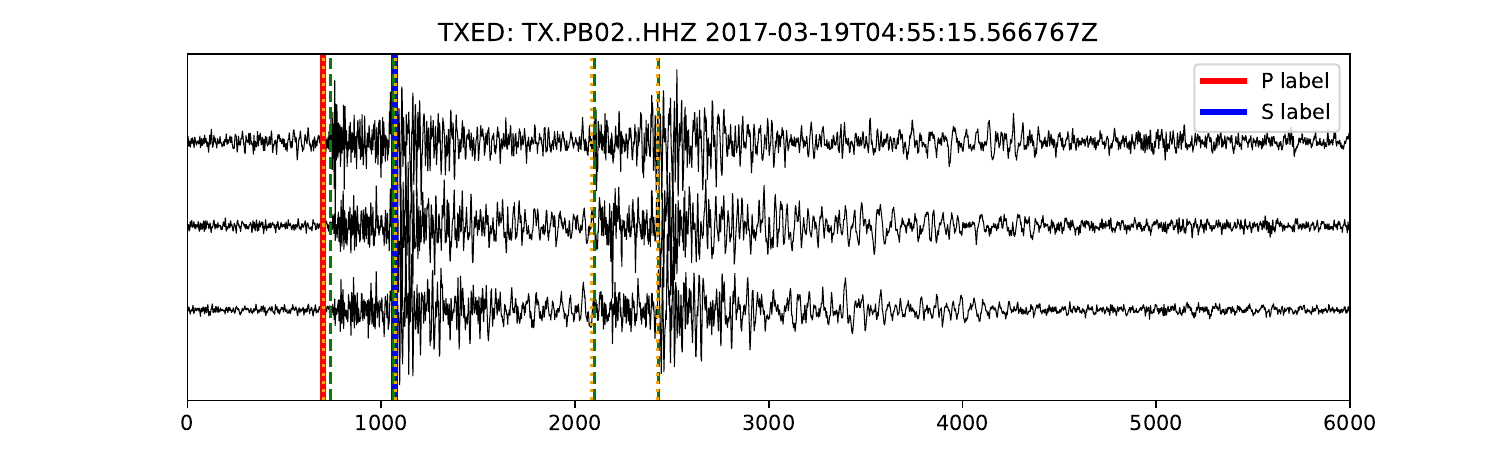}
        \end{subfigure}
        \hfill
        \begin{subfigure}[b]{0.5\textwidth}
        \centering
        \includegraphics[trim={2.5cm 0cm 2.5cm 0cm},clip,width=\textwidth]{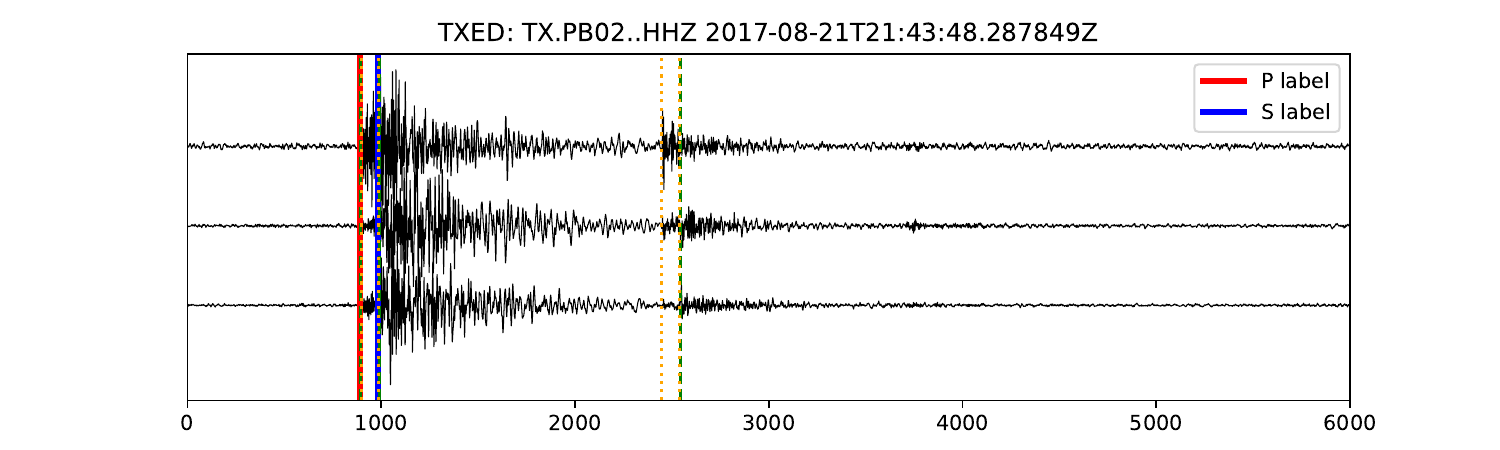}
        \end{subfigure}

        \begin{subfigure}[b]{0.5\textwidth}
        \centering
        \includegraphics[trim={2.5cm 0cm 2.5cm 0cm},clip,width=\textwidth]{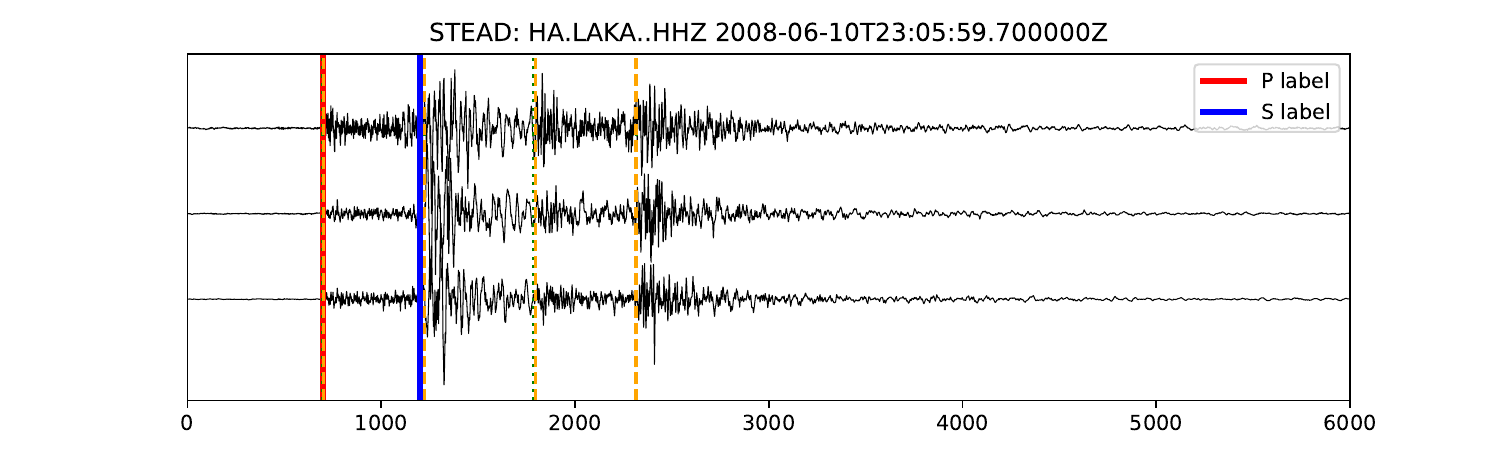}
        \end{subfigure}
        \hfill
        \begin{subfigure}[b]{0.5\textwidth}
        \centering
        \includegraphics[trim={2.5cm 0cm 2.5cm 0cm},clip,width=\textwidth]{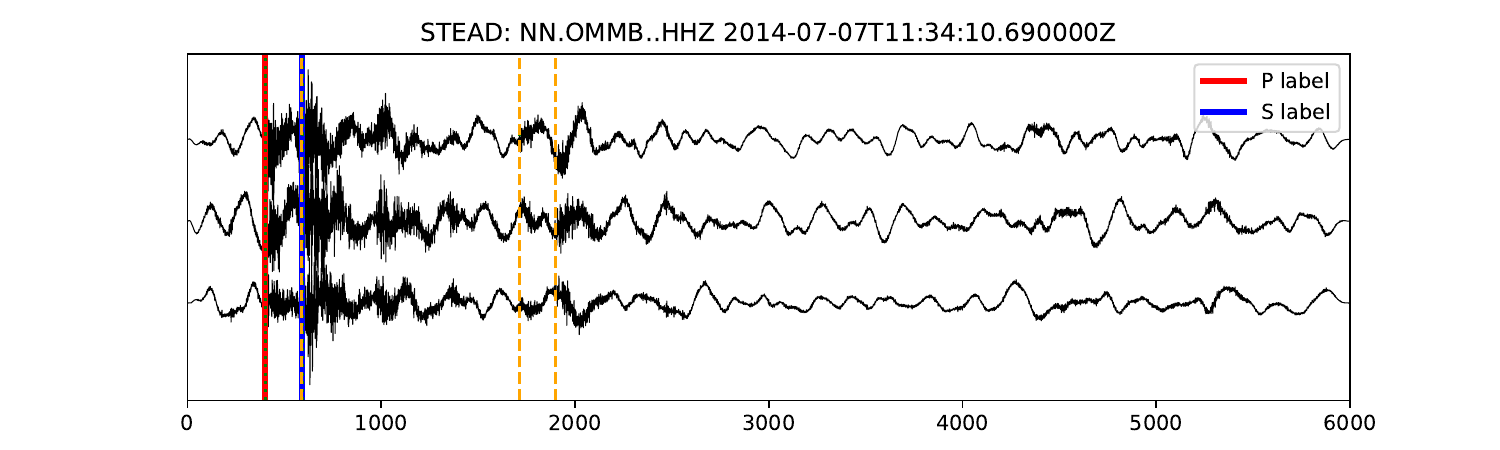}
        \end{subfigure}

        \begin{subfigure}[b]{0.5\textwidth}
        \centering
        \includegraphics[trim={2.5cm 0cm 2.5cm 0cm},clip,width=\textwidth]{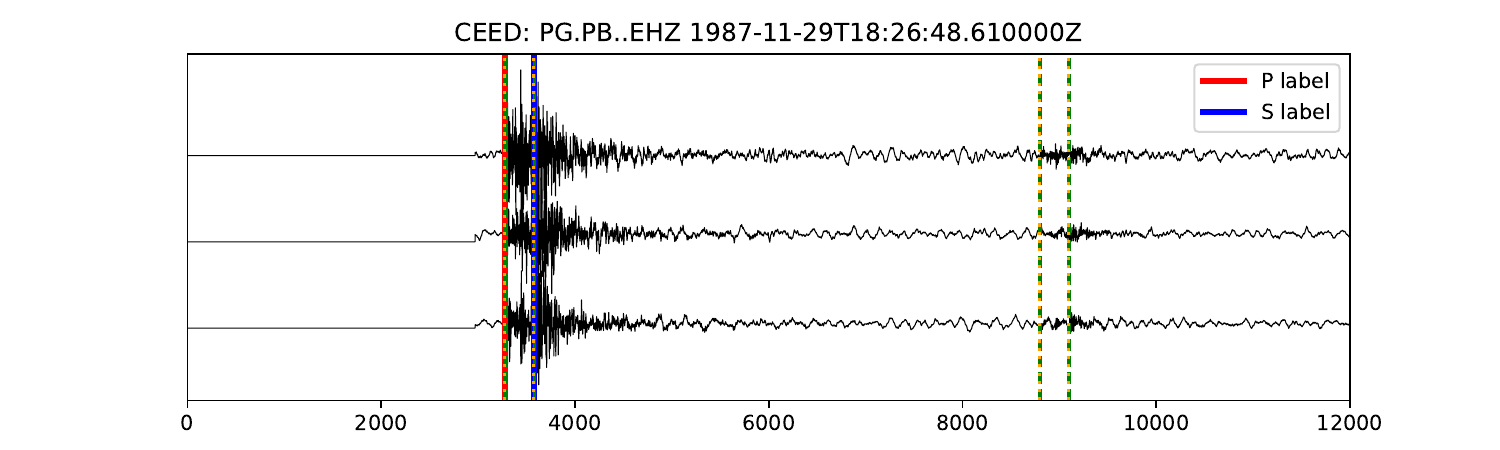}
        \end{subfigure}
        \hfill
        \begin{subfigure}[b]{0.5\textwidth}
        \centering
        \includegraphics[trim={2.5cm 0cm 2.5cm 0cm},clip,width=\textwidth]{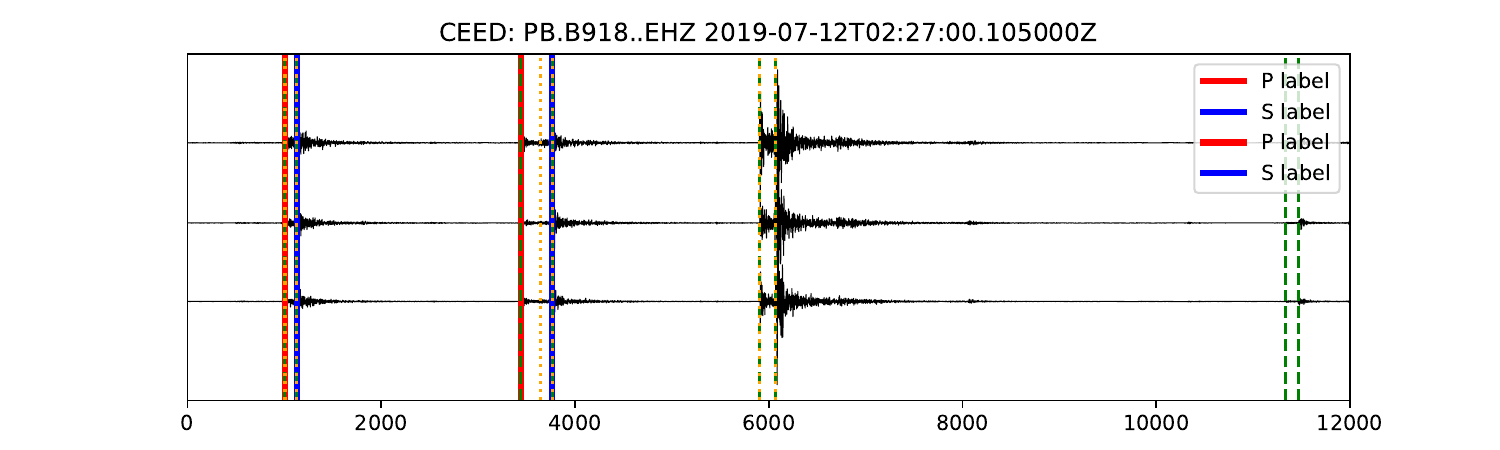}
        \end{subfigure}

        \caption{Samples that contain more earthquakes than those labeled. Dotted lines indicate modelensemble inferred arrivals.}
        \label{fig:multiples}
    \end{figure*}

    Despite the many benefits of deep learning accelerated research, there are unaddressed issues with the high volume of machine learning enabled algorithms \citep{KAPOOR2023100804,snakeoil} in the broader research community. Data sets across domains, spanning text, image, and audio, contain errors that degrade model performance and evaluation \citep{Northcutt2021}. Strategies have been devised to improve datasets and learn from them in the presence of faulty labels \citep{Cordeiro2020,Northcutt2021b} in the so-called Data-Centric AI approach \citep{oala2024dmlrdatacentricmachinelearning}. Data quality has been highlighted as a major factor determining model performance \citep{gunasekar2023textbooksneed}, as model performance gains from improved datasets outpace those from increasing model complexity. The goal of this paper is to improve the quality of seismological datasets for deep learning.

    As established phase picking models are applied to challenging settings, several studies have documented some shortcomings \citep{ruppert2022,ysp_2023,noelwest25,AguilarSuarez26} and solutions suggested \citep{park2025}.  Here, we analyze multiple data sets, including: the Stanford Earthquake Dataset (STEAD) \citep{stead}, the Iquique aftershocks dataset \citep{Woollam2019}, the Curated Pacific Northwest AI-ready seismic dataset (PNW) \citep{Ni2023}, the Italian seismic dataset for machine learning (INSTANCE) \citep{instance}, the Texas Earthquake Dataset for AI (TXED) \citep{texed}, the California Earthquake Event Dataset (CEED) \cite{zhu2025californiaearthquakedatasetmachine}, and the L'Aquila 2009 aftershocks dataset (AQ2009) \citep{aq2009}. This work was facilitated by the common access point to these datasets provided by Seisbench \citep{Woollam2022}. We analyzed the training, development, and test partitions of these datasets.

    We exclude datasets for which the waveforms are too short for available models, specifically the LenDB dataset of 27 seconds, for which padding or other signal modifications would be required to input the data into existing models, potentially harming model performance. We also avoided datasets with waveforms that are too long, like CREW \citep{AguilarSuarez_Beroza_2024}, for which only one model is available (SKYNET \cite{AguilarSuarez_Beroza_2025}). Following the same logic we exclude the MLAAPDE \citep{mlaapde} and NEIC \citep{neic} datasets, that cover  local, regional, and teleseismic distances, because their length does not reach the 5 minutes required to run SKYNET without modifications or padding. We exclude data from volcanic settings and ocean bottom seismometer recordings because there are not enough models to create a meaningful ensemble. We also exclude the SCEDC dataset due to its redundancy with the CEED dataset. 
    
    In a separate study (Aguilar Suarez \& Beroza, in prep) we document a number of shortcomings of ML-based earthquake catalogs. In an effort to understand their origin, we compared the phase picks used to construct ML catalogs to that that are used to create the USGS Comprehensive Catalog (ComCat). We found that about 10 percent of arrivals are missed by deep learning models, and as a result that up to 10 percent of known earthquakes in routine earthquake catalogs are missed in ML catalogs. This low recall rate is compounded by the occurrence of false positive detections. Both of these shortcomings motivate efforts to train more reliable phase picking models based on more reliable training datasets that are used to construct them.

\section{Errors}

    We take a Data-Centric approach, using established deep learning models PhaseNet \citep{Zhu2019} and EQTransformer \cite{EQTransformer} to flag potential label errors in the datasets by ensembling their outputs and comparing to the dataset labels. We recognize that this is an approximate approach in the sense that the models used for the evaluation were themselves trained on data with label errors and are not perfectly reliable indications of the presence, absence, or timing of phase arrivals.  Nevertheless, they have sufficient predictive value to identify large numbers of label errors.  We recognized four main categories of errors: (1) unlabeled earthquakes, (2) earthquakes in noise samples, (3) false earthquakes, and (4) inaccurately timed arrivals. 
    
    \subsection {Unlabeled Earthquakes} 
    The first and most prevalent category of error is the presence of earthquake arrivals for which there are no labels. That is, uncataloged earthquakes in the waveform data in addition to the ones that are labeled. Figure \ref{fig:multiples} shows a variety of examples from the analyzed datasets that contain more earthquakes than the labeled ones. These labels are faulty because they are incomplete. The uncataloged earthquakes have both low and high amplitude and are located before and after the labeled signal, such that we observed no systematics in the position or relative amplitude of the labeled vs. unlabeled events. Intuitively, small uncataloged earthquakes would follow larger cataloged ones, but during intense aftershock sequences there are both small and larger earthquakes occurring closely in time. The number of unlabeled earthquakes ranges from one to ten. Not surprisingly, the datasets with longer waveforms, like INSTANCE and CEED show a greater range.  Table \ref{tab:tab_1} summarizes error prevalence in the eight datasets examined. The error rate ranges from nearly zero to 8 \%. The dataset with the lowest prevalence of this type of faulty label is STEAD, followed by TXED, PNW, ETHZ, Iquique, CEED, AQ2009, and finally INSTANCE, with the highest prevalence of this error at 7.98 \%.

    \begin{table*}[ht!]
        \centering
        \begin{tabular}{l|r|r|r}
             \rowcolor{lightgray}
            \textbf{Dataset} & \textbf{Total Examples checked} & \textbf{Faulty examples} & \textbf{Prevalence} \\ 
             Iquique & 13,400  & 515 & 3.84 \%\\ 
             ETHZ & 36,743  & 949 & 2.58 \% \\
             PNW & 183,909  &  4,970 & 2.15 \% \\
             TXED & 312,231 & 4,672 & 1.49 \% \\
             STEAD & 1,030,231 & 149  & 0.01 \% \\
             INSTANCE & 1,159,249 &  92,561 &  7.98 \%\\
             AQ2009 & 1,258,006 &  87,008 & 6.92 \% \\
             CEED   &  4,140,530  &  253,309  & 6.12 \%

        \end{tabular}
        \caption{Summary numbers of the multiplet error rate.}
        \label{tab:tab_1}
    \end{table*}

    \subsection {Earthquakes in Noise Samples}
\begin{figure*}[]
        \begin{subfigure}[b]{0.5\textwidth}
        \centering
        \includegraphics[trim={2cm 0cm 2cm 0cm},clip,width=\textwidth]{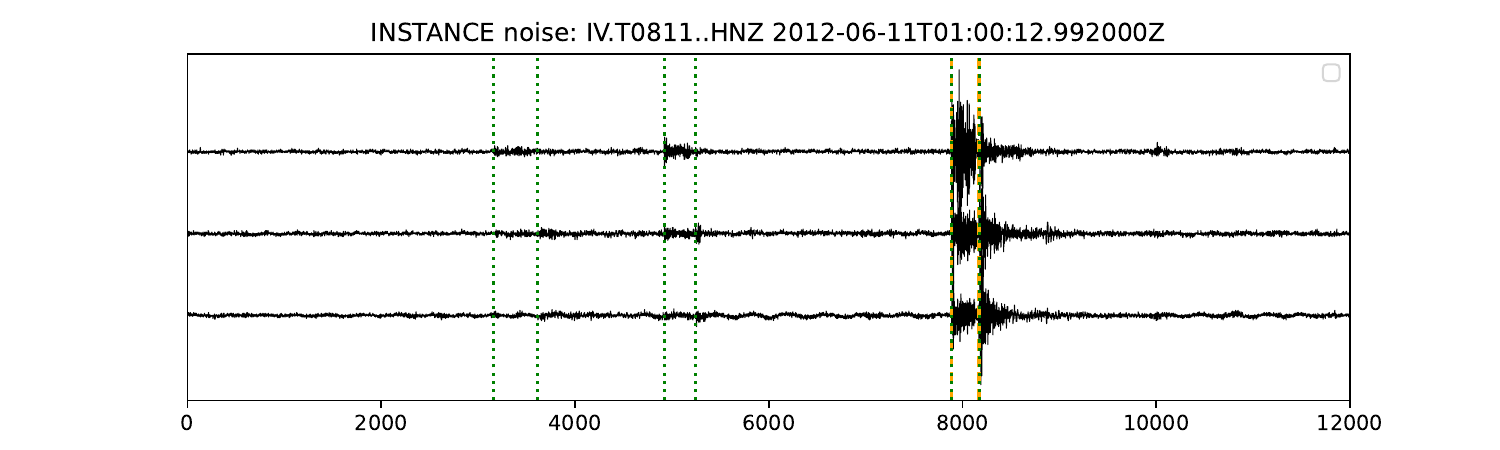}
        \end{subfigure}
        \begin{subfigure}[b]{0.5\textwidth}
        \centering
        \includegraphics[trim={2cm 0cm 2cm 0cm},clip,width=\textwidth]{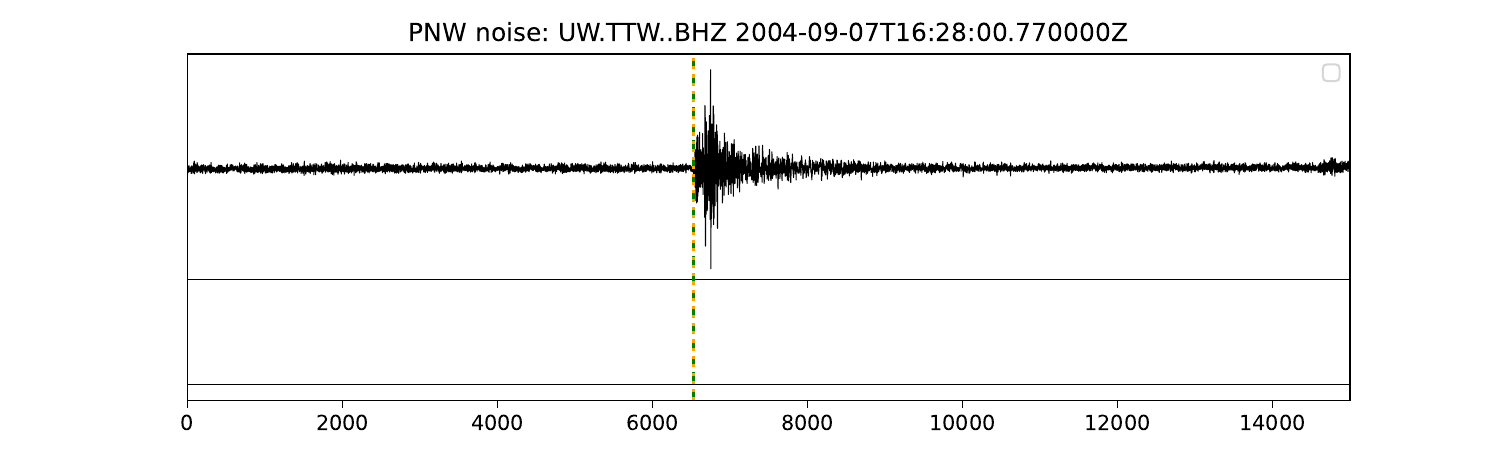}
        \end{subfigure}
        \begin{subfigure}[b]{0.5\textwidth}
        \centering
        \includegraphics[trim={2cm 0cm 2cm 0cm},clip,width=\textwidth]{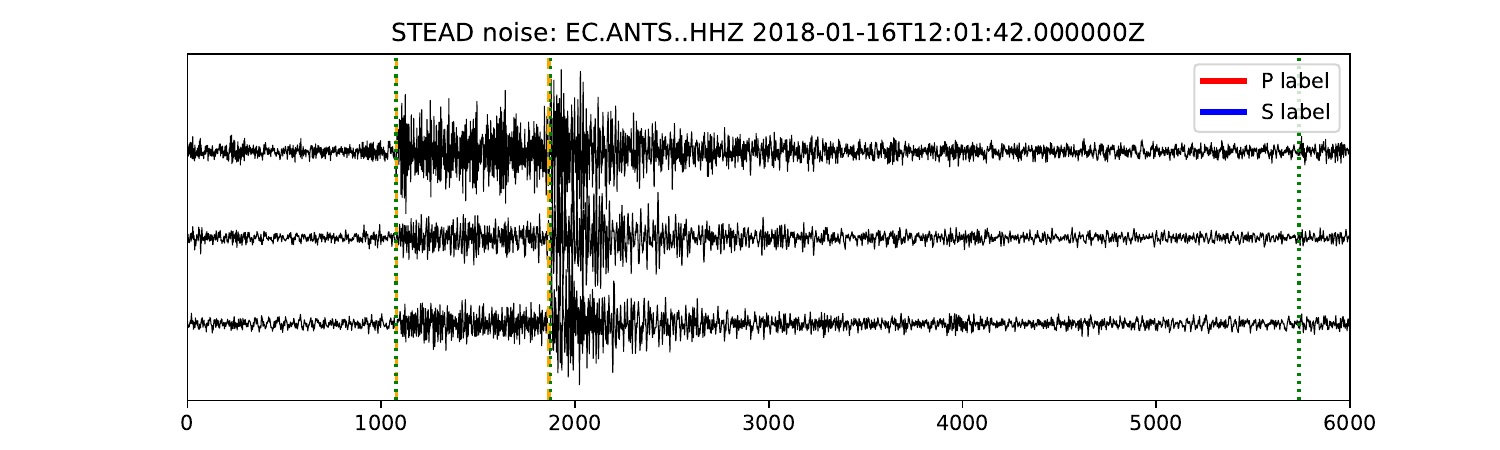}
        \end{subfigure}
        \begin{subfigure}[b]{0.5\textwidth}
        \centering
        \includegraphics[trim={2cm 0cm 2cm 0cm},clip,width=\textwidth]{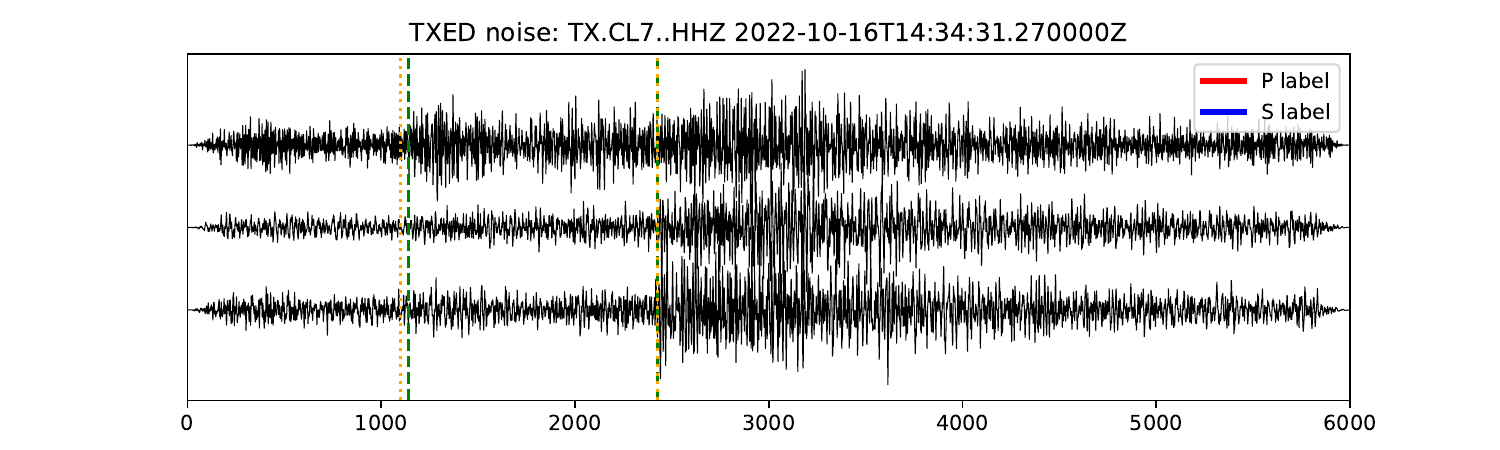}
        \end{subfigure}
        \caption{Noise examples for which earthquake arrivals were found. PhaseNet and EQTransformer arrivals are shown with the dotted lines.}
        \label{fig:noise_samples}
    \end{figure*}

    The second type of error is the presence of earthquake signals in samples that are labeled as noise, which can be considered false negatives. For this type of error we are restricted to the STEAD, TXED, INSTANCE, and PNW datasets, which contain noise samples, whereas the other datasets do not. In this case if our ensemble found and agreed on at least one arrival, the example was targeted for further analysis. Figure \ref{fig:noise_samples} shows some of those examples that are in the datasets as noise labels, but in fact do contain earthquake waveforms, from one or more events. STEAD contains 235,426 noise samples, from which we found arrivals in 208. After manual inspection we removed 14 samples because they are glitches rather than earthquake arrivals.  The prevalence of this error modality in the INSTANCE dataset was estimated to be 1.92 \%, with 2549 examples like this one out of the 132,288 total noise examples. Out of the 51,462 noise samples of the PNW dataset, 446 were found to contain earthquake arrivals, making up 0.87 \% of the noise partition. Finally, out of the 207,458 noise samples of the TXED dataset, 458 were flagged. In contrast to the other datasets, many of these flagged samples do not contain clear earthquake signals, and we had to manually check each one, resulting in only 16 samples (less than 0.01 \%) with clear earthquake signals. Table \ref{tab:tab_noise} summarizes the number of samples analyzed, alongside the error prevalence in the noise datasets.

    \begin{table*}[h!]
        \centering
        \begin{tabular}{l|r|r|r}
             
           \rowcolor{lightgray} \textbf{Noise Dataset} & \textbf{Total Examples} & \textbf{Faulty examples} & \textbf{Prevalence} \\  
              STEAD & 235,426 & 194 & 0.09 \%\\
             TXED & 207,458 & 16 & 0.008 \%\\
              INSTANCE & 132,288 & 2,549 & 1.92 \%\\
              PNW & 51,462 & 446 & 0.87 \%

        \end{tabular}
        \caption{Prevalence of errors in noise examples that contain unlabeled earthquakes.}
        \label{tab:tab_noise}
    \end{table*}

    \subsection {False Earthquakes}
\begin{figure*}[]

        \begin{subfigure}[b]{0.5\textwidth}
        \centering
        \includegraphics[trim={2cm 0cm 2cm 0cm},clip,width=\textwidth]{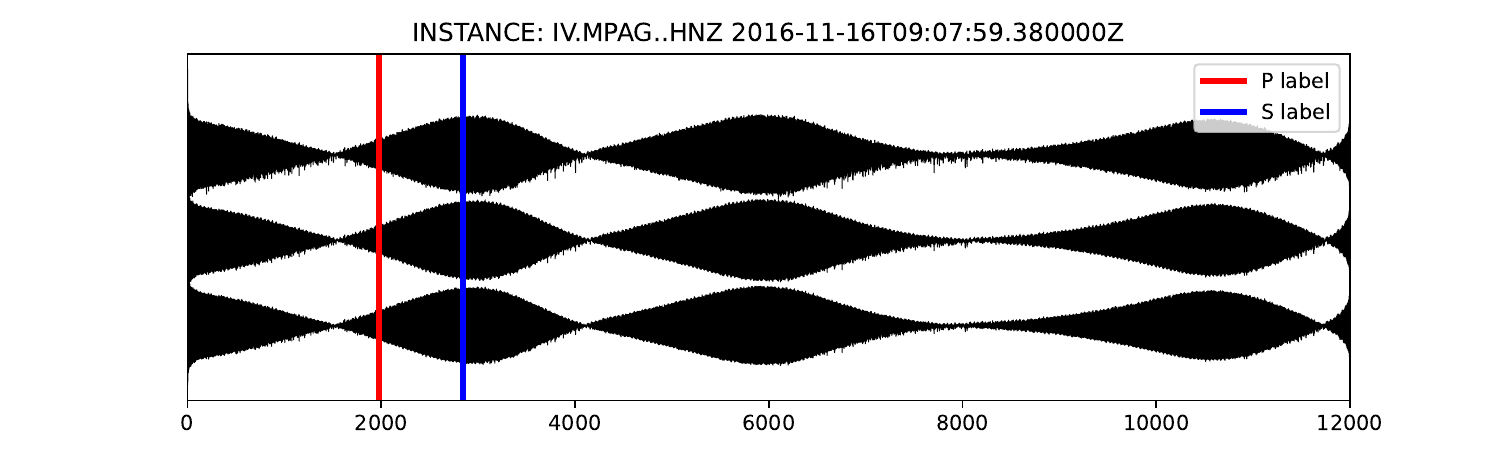}
        \end{subfigure}
        \begin{subfigure}[b]{0.5\textwidth}
        \centering
        \includegraphics[trim={2cm 0cm 2cm 0cm},clip,width=\textwidth]{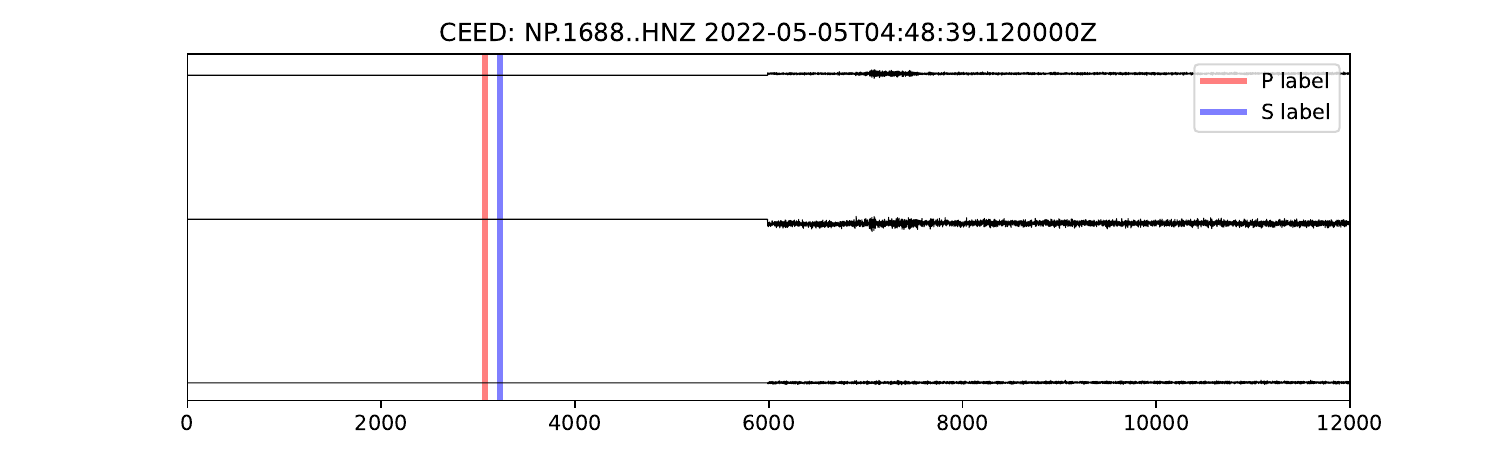}
        \end{subfigure}
        \caption{Labeled arrivals for which there are no earthquake waveforms.}
        \label{fig:fp_samples}
    \end{figure*}

    The third type of error we identified is for samples that contain labeled arrivals, but the waveforms do not show earthquake signals because the signal is absent or incomplete, such that these can be considered as false positives. We initially recognized this failure mode by using a low threshold of 0.2 for phase picking. Any example for which PhaseNet and EQTransformer predicted no picks with the low threshold of 0.2 was flagged. These types of samples come mostly from accelerometer data, as shown in Figure \ref{fig:fp_samples}. For the sample on the right panel of Figure \ref{fig:fp_samples}, the arrivals are in a section of the seismogram that has constant amplitude, most likely due to an instrument outage, some form of padding during postprocessing, and/or an archival artifact. Due to our filtering criteria not translating well from one dataset to another, and a tendency to flag single channel data excessively, we did not quantify the prevalence of this type of error.

    \begin{figure*}
    \begin{subfigure}[b]{0.5\textwidth}
    \centering
    \includegraphics[trim={2cm 0cm 2cm 0cm},clip,width=\textwidth]{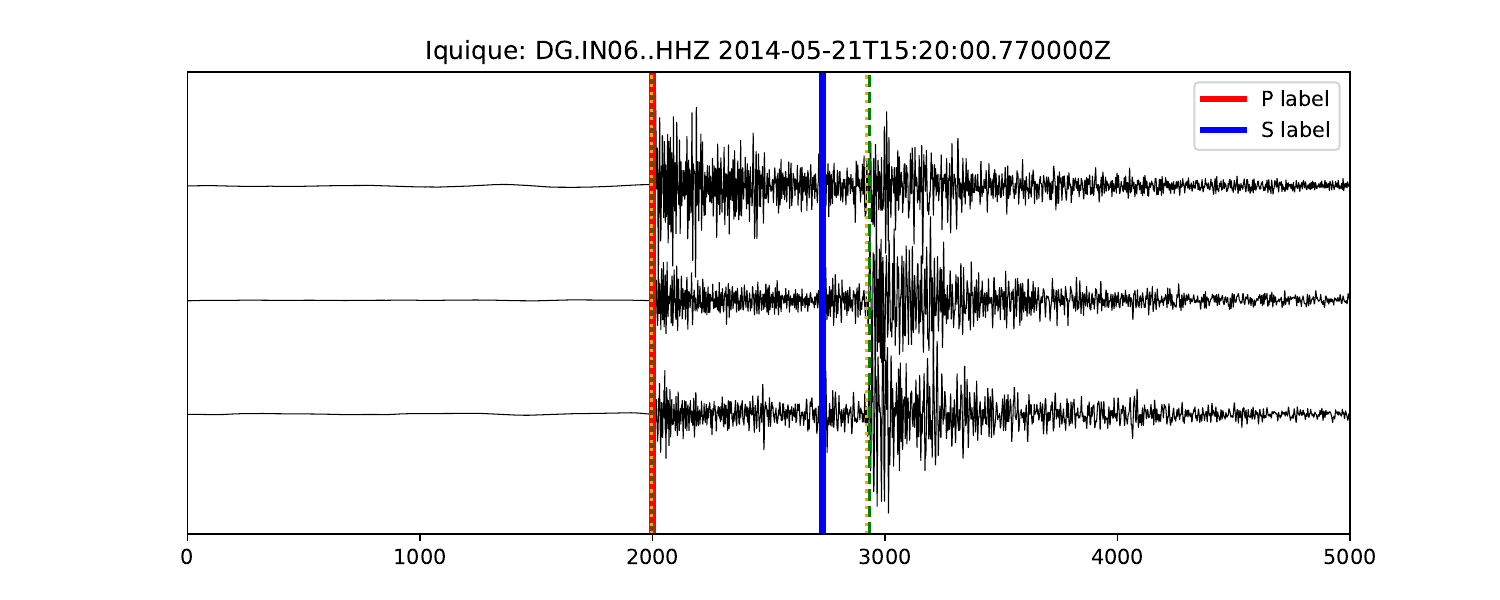}
    \end{subfigure}
    \begin{subfigure}[b]{0.5\textwidth}
    \centering
    \includegraphics[trim={2cm 0cm 2cm 0cm},clip,width=\textwidth]{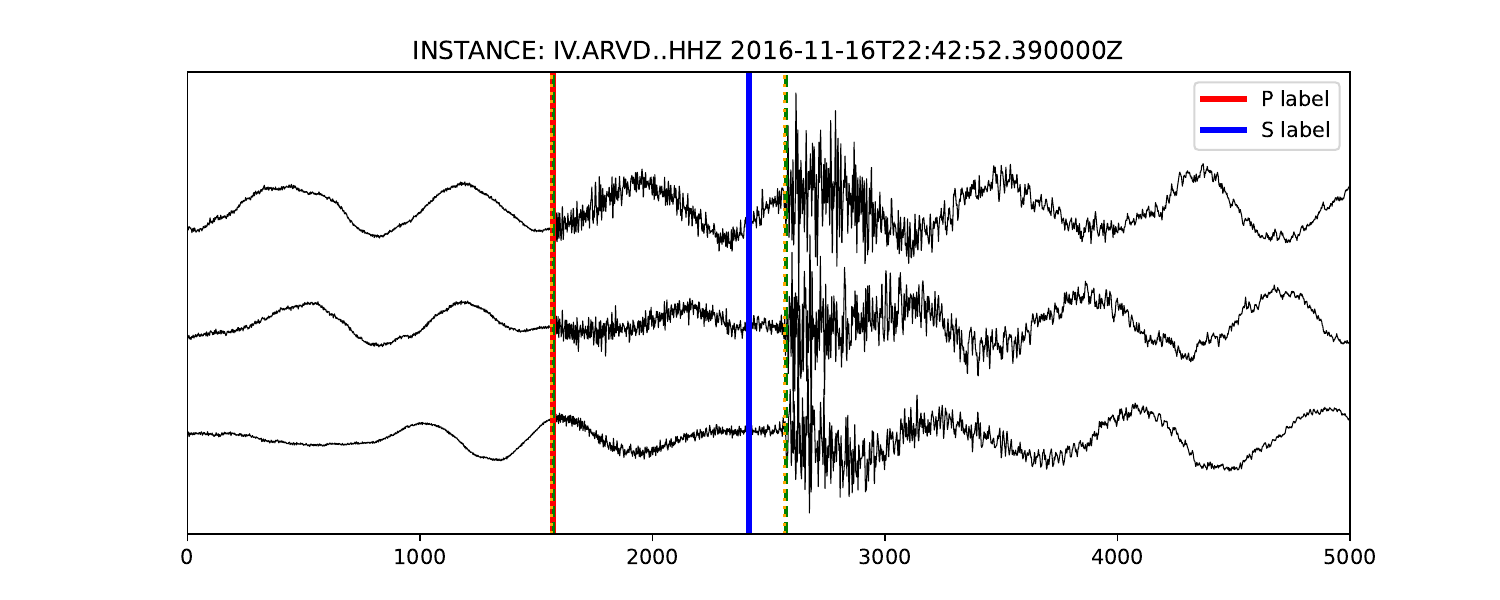} 
    \end{subfigure}

    \caption{Samples with inaccurately timed arrivals. Dotted lines show the machine learning ensemble. The name of the dataset is indicated on the title of each panel.}
    \label{fig:mistimed}
    \end{figure*}

    \subsection {Inaccurate Timing}
    The fourth type of error is labeled arrivals with inaccurate timing, that is, P or S labels that are off by several seconds from their actual arrival times in the waveforms. We compared the labeled arrivals with the predictions from the deep learning ensemble, and marked those with a discrepancy larger than 2 seconds for further analysis. Figure \ref{fig:mistimed}(left) shows an example of a badly timed S arrival, which disagrees with the ensemble by several seconds, and is clearly labeled as arriving a couple of seconds before the onset of the S wave energy. We identified only a few such examples, but due to the process of automating the screening not easily translating from one dataset to another, we decided to leave this detailed analysis for future research.


    \section{From Big Data to Good Data}
    The findings outlined here are intended to lead to improvements in performance for deep learning models, which will have positive effects on downstream tasks, such as phase association and location, and will improve the reliability of automated earthquake catalogs. \cite{Northcutt2021} estimated an average error rate of 3.3 \% for the labels in the test sets of the 10 most widely used datasets for image, text and audio for machine learning. Our study is most likely a lower bound of the error rate, which is on average 3.9 \% across all the datasets analyzed. This number does not include all recognized types of errors, but only those for which we could create an automated, trustworthy, and translatable procedure.

    In our approach, we noticed that PhaseNet declares many more arrivals than EQTransformer.  We used the conservative version of EQTransformer for our test.  We observed several examples that do contain multiple earthquakes, but only PhaseNet and not EQTransformer identified them, so we do not treat them as errors of the multiplet type based on the ensemble. Thus, we believe the numbers presented here should be taken as a lower bound of the error prevalence.

    The samples with multiple arrivals from multiple events highlight the complexity of earthquake sequences, with numerous earthquakes occurring closely in time. Specifically for the INSTANCE samples in Figure \ref{fig:multiples} in the 120 s long waveforms there can be up to 10 earthquakes in quick succession, for which accurate labeling is a challenge, even for state of the art machine learning pickers. We also note that the models we are using for evaluation were themselves trained with data that contains label inaccuracies. The AQ2009 and CEED datasets attempt to work around this shortcoming by allowing the dataset to contain labeled arrivals from more than one single earthquake, but they still fall short during intense seismicity sequences (Figure \ref{fig:multiples} bottom right). Models can be trained to deal with such dense earthquake signals with appropriate data augmentation \citep{zhuaugmentation}, as exemplified in Figure 2 of \cite{AguilarSuarez_Beroza_2025} and Figure S3 of \cite{tan2025}. While we found no systematics in the relative amplitude or positions of the labeled and unlabeled earthquakes, we suspect we are missing samples for which the amplitude contrast is extreme, that is a very low amplitude signal next to a very high amplitude signal. 
    
    Errors due to unlabeled earthquakes are an order of magnitude more prevalent than errors due to faulty noise samples (3.8 \% vs 0.7 \%), as detailed in Tables \ref{tab:tab_1} and \ref{tab:tab_noise}. The multiplet error rate is lower for STEAD and TXED, which are the datasets with the shortest waveforms, at 60 seconds each. This contrasts with the other datasets, which contain 120 or 150 second long waveforms, which can inherently contain more unlabeled earthquakes. This might also be behind the fact that both STEAD and TXED have the lowest prevalence of earthquake signals in samples labeled as noise. We believe, however, that the low prevalence of error in STEAD is a consequence of the careful quality control performed while assembling that dataset as reported by \cite{stead}, who estimated the error rate to be less than 1 \%. Our findings suggest that the error rate might be dramatically lower, closer to 0.02 \%.

    The errors found in this study likely include the biases of the data used to train both PhaseNet and EQTansformer, which will limit their ability to detect all errors comprehensively. We identified a number of inconsistencies with more regional earthquake waveforms, which are not well represented in STEAD or the Northern California data used to train PhaseNet. We manually checked random samples of flagged data and adjusted our selection criteria in order to avoid valid data being declared as faulty. There are many examples for which the arrivals are visible but neither PhaseNet nor EQTransformer declare a pick for all labeled phases, such that there is only partial quality control performed on such samples.

    One of the reasons for the prevalence of errors is that the process of assembling these datasets begins with gathering metadata from available catalogs. The corollary is that catalogs assembled on a routine basis have a variety of shortcomings, including but not limited to, bad measurements of arrival times, and mis-association of false picks. On a event-by-event basis, not all visible waveforms are picked, especially when analysts are dealing with intense seismicity sequences.



    \section{Future Directions}

    We documented the pervasive nature of label errors across seismological machine learning datasets, and quantified it for the two most prevalent types.  We find an average error rate of 3.9 \%. While the automated nature of our work allowed us to process several million waveforms, selected data was visually inspected, which suggests that we are not yet ready for full automation. Our approach has some limitations due to the nature of the ensemble, which can be biased to agree with data it was trained with. Nevertheless, the results shown here testify to the utility of our approach.
    The data that has been declared as faulty can be ignored, or potentially fixed, which might entail a combination of semisupervised learning and careful  manual supervision. The results outlined here aim to improve the quality of seismological machine learning datasets, which will in turn improve the performance of models trained on those datasets. 

    We have focused on benchmarks for seismic phase picking, but downstream tasks would benefit from the creation of additional benchmarks for such tasks as catalog building \citep{linville2019},  phase association \citep{pennington25,Puente2025} , and event location \citep{yu2025accuracy}.



		\begin{acknowledgements}
			The authors thank two anonymous reviewers and the editors for helping improve this manuscript. This work was supported by the Department of Energy (Basic
Energy Sciences; Award Number DE-SC0026319)
		\end{acknowledgements}
		
		\section*{Data and code availability}
        Files containing the names of the examples in each dataset, alongside the reasons for their selection are available here: \textbf{https://github.com/albertleonardo/labelerrors/}
        We are in the process of integrating a new metadata field to Seisbench such that the examples that should be avoided can be removed from training or testing routines.

		\section*{Competing interests}
		The authors declare no competing interests.\newline This is an LLM free manuscript.

        \bibliography{mybibfile}
		
	\end{document}